**Stochastic Dynamics in Quenched-in Disorder and Hysteresis**

Giorgio Bertotti, Vittorio Basso, and Alessandro Magni

IEN Galileo Ferraris, INFM, Corso M. d'Azeglio 42, I-10125 Torino, Italy

**Abstract**

The conditions under which relaxation dynamics in the presence of quenched-in disorder lead to rate-independent hysteresis are discussed. The calculation of average hysteresis branches is reduced to the solution of the level-crossing problem for the stochastic field describing quenched-in disorder. Closed analytical solutions are derived for the case where the disorder is characterized by Wiener-Lévy statistics. This case is shown to be equivalent to the Preisach model and the associated Preisach distribution is explicitly derived, as a function of the parameters describing the original dynamic problem.

**Introduction**

There are many physical situations in which thermodynamic equilibrium is difficult to reach. The system may be trapped in long-living metastable states, and over ordinary time scales it may appear as permanently occupying a non-equilibrium state dependent on past history. Under these conditions, the relevant interpretative scheme is rate-independent hysteresis rather than equilibrium thermodynamics. In magnetism, micromagnetics and Landau-Lifshitz-Gilbert equations provide in principle the theoretical framework for the description of magnetization processes. Magnetic hysteresis should emerge from this description after appropriate averages over structural disorder. In its most general form, this problem is of discouraging difficulty. In this paper, we limit the discussion to a much simpler situation, that of a scalar system governed by relaxation dynamics of the form

$$\gamma \frac{dX}{dt} = H(t) - H_F(X) \tag{1}$$

where $\gamma > 0$ is the friction constant, $H(t)$ is the time-varying driving field and $X(t)$ is the system response. The internal structure of the system is described by the free energy gradient $H_F(X)$. Due to structural disorder, $H_F(X)$ will fluctuate at random as a function of $X$.

Néel [1] was the first to develop the concept of random energy landscape, in the framework of an explanation of the Rayleigh law on a statistical basis. Kronmuller and Reininger [2] started from this concept, investigating the properties of the derivative of the energy profile $H_F(X)$ on the basis of the local fluctuations of the defect concentration, supposed to act as pinning centers. In our approach we will not enter in any detail concerning the physical origin of the fluctuations in the free energy gradient.

We shall discuss the problem in the limit of low temperatures, where thermal agitation can be neglected and $H(t)$ only is the relevant driving force.

The hysteresis properties of the model emerge from the study of the trajectories $H(X)$ of the system in the $(X,H)$ plane. In particular, one finds that the rate-independent trajectories obtained in the limit $|dH/dt| \to 0$ are given by the sequence of points at which $H_F(X)$ reaches progressively higher field levels for the first time. On this basis, one can reduce the problem of the calculation of the average "magnetization curve" $X(H)$ of the system to the solution of the level-crossing problem for the stochastic process $H_F(X)$. By exploiting the mathematical results known in this field, we shall obtain an analytical solution for $X(H)$ in the case where quenched-in disorder follows Wiener-Lévy statistics. Remarkably, this analytical solution turns out to be equivalent to Preisach-type hysteresis. This result represents a limited but important example of derivation of macroscopic hysteresis properties from microscopic, non-equilibrium dynamic considerations.

**Hysteresis and the level-crossing problem**

The basic aspects of the dynamics described by Eq.(1) are made clear by considering the phase portrait of Eq.(1) in the $(X,H)$ plane when the field increases at a given constant rate $dH/dt$ (Fig. 1). Let us consider in particular the rate-independent limit $dH/dt \to 0$. In this limit, the trajectory obtained by integrating Eq. (1) is made of a sequence of stable states characterized by $H = H_F(X)$, connected by horizontal unstable branches (Barkhausen jumps) where $dH/dX = 0$ and $dX/dt = (1/\gamma)(H - H_F(X))$. This description can be recast in the following more general form, particularly useful when $H_F(X)$ is a stochastic process: the rate-independent trajectory under increasing field consists of the sequence of points $H = H_F(X)$ at which $H_F(X)$ reaches progressively higher field levels $H$ for the first time. This creates a direct connection between the trajectory and the statistics of the *level-crossing problem* for the stochastic process $H_F(X)$.

Let us restrict the analysis to the case where $H_F(X)$ is a Markovian stochastic process. This means that the evolution of the process under given initial conditions, e.g., $H_F = H_0$ at $X = X_0$, depends on these conditions only and not on the details of the process for $X < X_0$. Let us consider the process in the interval $(H_-, H_+)$, with $H_- < H_0 < H_+$, as shown in Fig. 2. Level-crossing is studied by following the process in its diffusion away from $H_0$, until it reaches one of the levels $H_-$ or $H_+$ for the first time. Here the process is stopped and removed. Mathematically, this feature is dealt with by introducing absorbing boundary conditions at $H_F = H_-$ and $H_F = H_+$. The case relevant to the study of hysteresis is the one where a realization of the process starting at $X = X_-$ arbitrarily close to the lower boundary $H_-$ but *not* absorbed by it, crosses the upper boundary $H_+$ at $X = X_+$, with $X_+ > X_-$ (see Fig. 2b). The crossing value $X_+$ is a random variable, described by a certain conditional probability distribution, normalized to unity, which we denote by $p_{LC}(X_+, H_+ \mid X_-, H_-)$. There are methods discussed in the literature to derive this distribution [3, 4]. For the moment, we shall simply assume that it has been calculated and is known for a given process.

Let us discuss how these results apply to the problem of hysteresis. First of all, notice that the system of Fig.1 obeys the property known as *return-point memory*. This property can be defined as follows. Suppose $H$ is decreased from large positive values down to $H_1$, then it is increased up to $H_2$ and finally it is decreased again. If return-point memory holds, when $H$ reaches again the value $H_1$ the system returns back to the exact state it occupied when the field $H_1$ was reached for the first time. It has been proven [5] that return-point memory holds for the individual trajectories of Fig. 1.

Suppose we consider the trajectory of Fig. 1 when the field is decreased from some large initial positive value ( positive saturation ) down to $H = H_1$, where it is reversed and increased. Let $X_1$ be the value of $X$ at the reversal point. The distribution $p_+(X, H; H_1)$ of the system response $X$ after the reversal is controlled by the level-crossing probability $p_{LC}(X, H \mid X_1, H_1)$ (see Fig. 2b, with $H_+ = H$, $H_- = H_1$) and by the distribution $p_1(X_1; H_1)$ of the reversal value $X_1$, according to the expression

$$p_+(X, H; H_1) = \int_{-\infty}^{X} p_{LC}(X, H \mid X_1, H_1) p_1(X_1; H_1) \, dX_1; \qquad H > H_1 \qquad (2)$$

Let us now suppose that the field is increased up to $H = H_2$, $X = X_2$, and then reversed again. Let $p_2(X_2; H_2, H_1)$ be the distribution of $X$ at the new reversal point. $p_2(X_2; H_2, H_1)$ coincides with $p_+(X_2, H_2; H_1)$ and can be calculated from Eq.(2). Each point $(X, H)$ of the new descending branch is a potential upward reversal point. Therefore, because of return point memory, it is governed by an equation identical to Eq. (2), in the field interval $(H, H_2)$. This means that the distribution $p_-(X,H;H_2,H_1)$ associated with the descending branch will be the solution of the inverse problem

$$p_2(X_2;H_2,H_1) = \int_{-\infty}^{X_2} p_{LC}(X_2,H_2 \mid X,H) p_-(X,H;H_2,H_1) dX; \qquad H_1 < H < H_2 \qquad (3)$$

This procedure can be continued to obtain $p_3(X_3;H_3,H_2,H_1) \equiv p_-(X_3,H_3;H_2,H_1)$, and then $p_+(X,H;H_3,H_2,H_1)$ and so on. Any desired higher-order reversal branch can be generated and studied this way. The average system response along the generic $n$-th order branch will be

$$X_\pm(H;H_n,...,H_2,H_1) = \int_{-\infty}^{+\infty} X \, p_\pm(X,H;H_n,...,H_2,H_1) dX \qquad (4)$$

where the "+" ("-") subscript denotes an ascending (descending) branch. We see that in general the higher-order branches will depend explicitly on the entire past sequence of reversal fields.

**Wiener-Lévy statistics**

Let us apply the previous analysis to the particular case where the free energy gradient $H_F(X)$ has the form

$$H_F(X) = \frac{X}{\chi} + H_p(X) \qquad (5)$$

In Eq. (5), the term $X/\chi$ represents some parabolic potential well ensuring the large-scale stability of the system, and $H_p(X)$, the pinning field, is a random process with zero mean, which we assume to coincide with the Wiener-Lévy process. In particular, this means that $<|dH_p|^2> = 2AdX$, with $A > 0$. Pinning fields with Wiener-Lévy statistics are of definite physical interest. They were experimentally observed in ferromagnetic single crystals [6, 7], and their introduction leads to a satisfactory general interpretation of the Barkhausen effect [8].

The solution of the level-crossing problem for the Wiener-Lévy process with drift (Eq. (5)) is known [4, 9]. One finds that Eq. (4) takes the simple analytical form

$$X_{\pm}(H;H_n,...,H_2,H_1) = \langle X_n \rangle \pm 2A\chi^2 \left[ \frac{H-H_n}{2A\chi} \operatorname{cth}\left(\frac{H-H_n}{2A\chi}\right) - 1 \right] \quad (6)$$

A remarkable property of this solution is that it can be mapped onto the Preisach model of hysteresis. This derives from the fact that Eq.(6) obeys return point memory, and another property, known as *congruency*, which represent the necessary and sufficient conditions for the description of a given system by the Preisach model [9, 10]. The congruency property can be defined as follows. In a system obeying this property, reversal branches starting from identical reversal fields are geometrically congruent, whatever the past field history. Congruency does not hold in general for a system described by Eqs. (2)-(4). However, it holds in the particular case of Wiener-Lévy statistics. In Eq.(6), in fact, the influence of past history is all lumped in the average value $\langle X_n \rangle$ at the last reversal point. Therefore, all reversal branches have the same geometrical shape and merely differ by a shift, controlled by the values of $H_n$ and $\langle X_n \rangle$. This proves the congruency property.

The conclusion is that there exists a Preisach distribution $p(h_c, h_u)$ (with $h_c$ and $h_u$ representing local coercive fields and local interaction fields, respectively) that describes the average hysteresis properties of the dynamics in Wiener-Lévy-type disorder. This distribution can be determined by first noting that, according to Eq. (6), any generic hysteresis branch depends on the difference $H-H_n$ only. It is known that, when this is the case, the Preisach distribution will depend on the local coercive field $h_c$ only. The dependence on $h_c$ is then obtained by calculating from Eq. (6) the derivative $d^2X_+/dH^2 = (\chi/2)\, p[(H-H_n)/2]$. The result is [11]:

$$p(h_c) = \frac{2}{A\chi} \frac{x\operatorname{cth} x - 1}{\operatorname{sh}^2 x}, \qquad x = \frac{h_c}{A\chi} \quad (7)$$

$p(h_c)$ is normalized to unity, $\int p(h_c)dh_c = 1$, and can be interpreted as a distribution of local pinning fields. Figure 3 shows a set of loops of increasing amplitude calculated from Eq.(6), or, equivalently, from standard Preisach integrals based on Eq.(7). The analytical derivation was tested by generating the same loops also through computer simulations. Individual random trajectories of the type shown in Fig. 1 were obtained from individual realizations of the process $H_F(X)$ and were averaged over many subsequent realizations. The results of the simulations differed by less than 1% from the analytic results of Fig. 3.

**Conclusions**

We have seen that, whatever the statistics of the pinning field of Eq. (5), return-point memory will hold under all circumstances. Congruency, on the other hand, was obtained only in the particular case of Wiener-Lévy statistics, and in general will fail for other types of random fields. This raises the interesting question of what type of property could play the role of congruency under more general conditions. Answering this question would show the direction in which one should generalize the Preisach model in order to approach a universal description of hysteresis and would give a natural way to introduce new classes of hysteresis models.

**Acknowledgments**

The authors are deeply indebted to Isaak Mayergoyz for illuminating suggestions about the general significance of the approach here proposed.

pertaining to a closed loop. Therefore, the Preisach distribution obtained in that case is incorrect.

**Figure Captions**

Fig. 1 – Free energy gradient $H_F(X)$ and corresponding individual trajectory $H(X)$ in the rate-independent limit $|dH/dt|\to 0$. The state $(X_1,H_1)$ is an upward reversal point.

Fig. 2 – Schematic representation of the level-crossing problem in the field interval $(H_-,H_+)$, starting from a given initial condition $(X_0,H_0)$ **(a)**, and in the limit $H_0\to H_-$ **(b)**.

Fig. 3 – Hysteresis loops calculated from Eq.(6). In the inset is shown the Preisach distribution $p(h_c)$ (Eq.(7)).

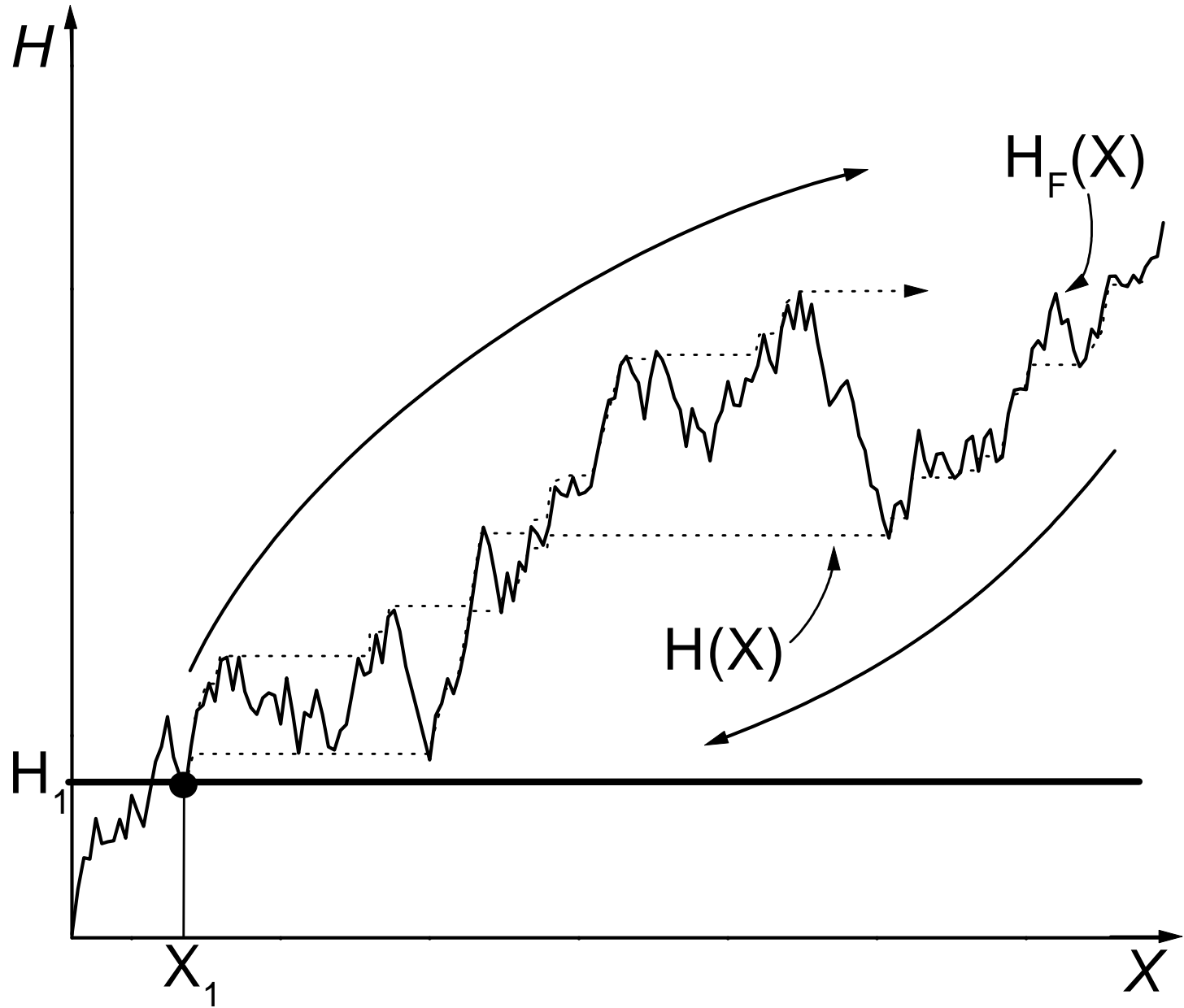

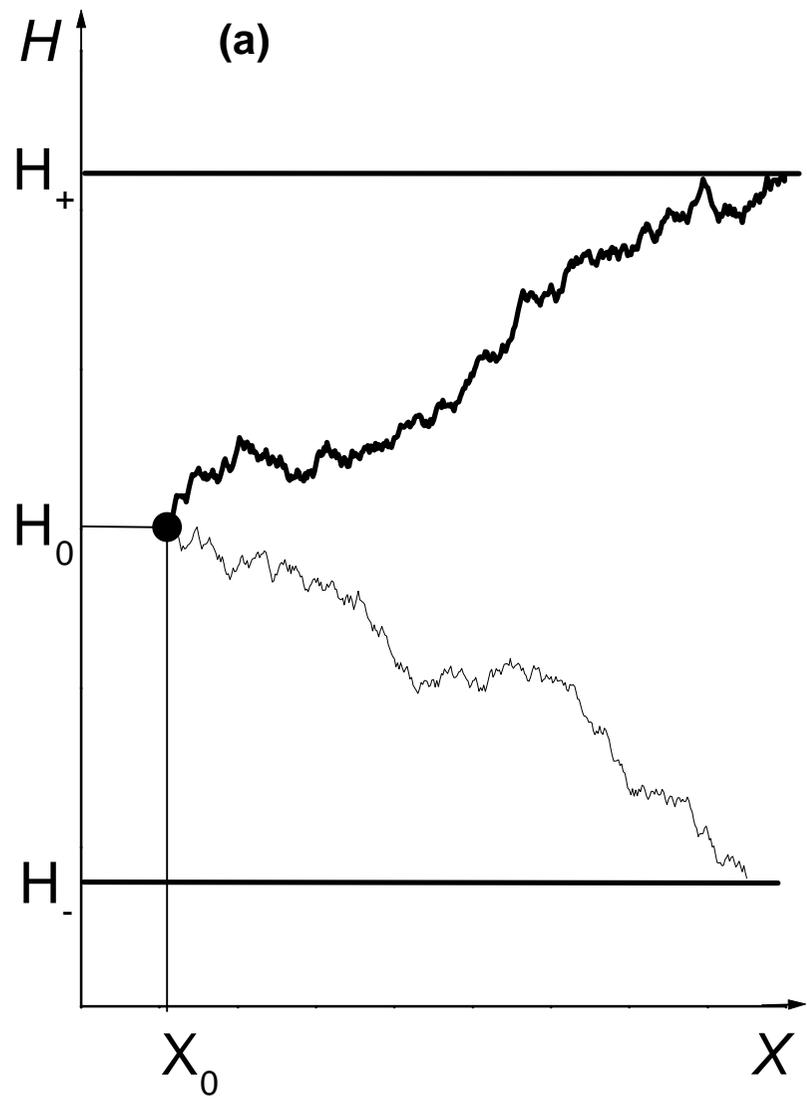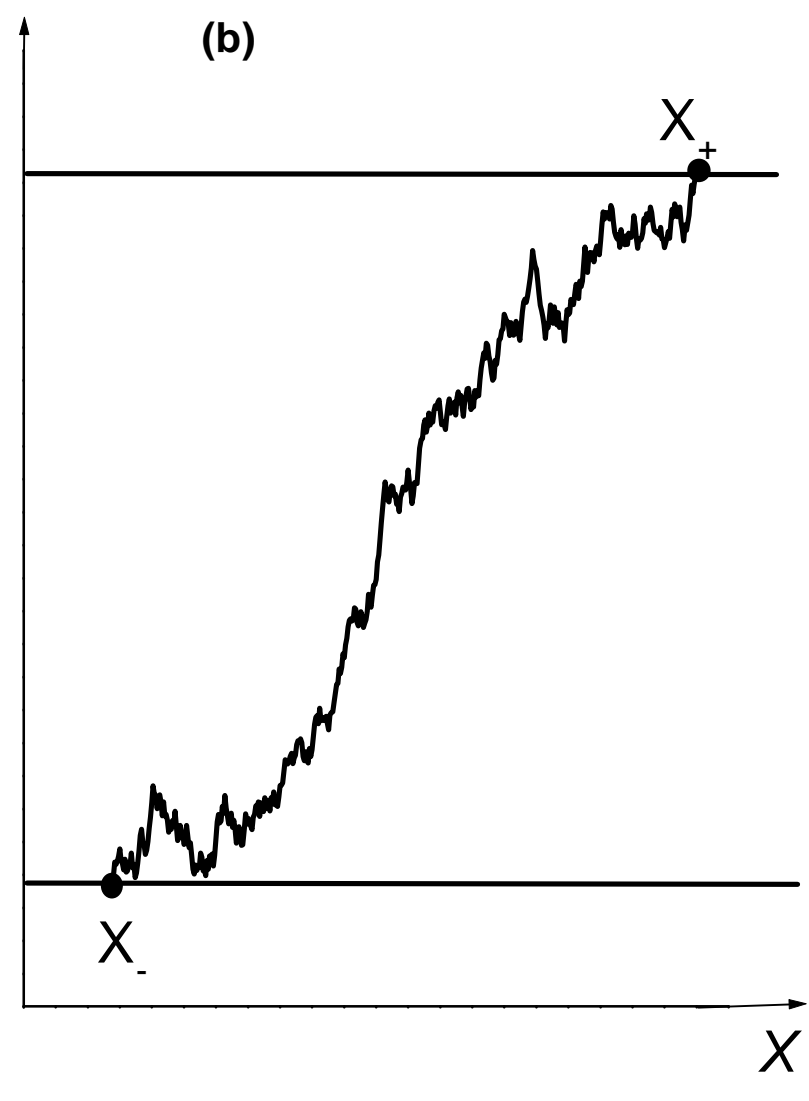

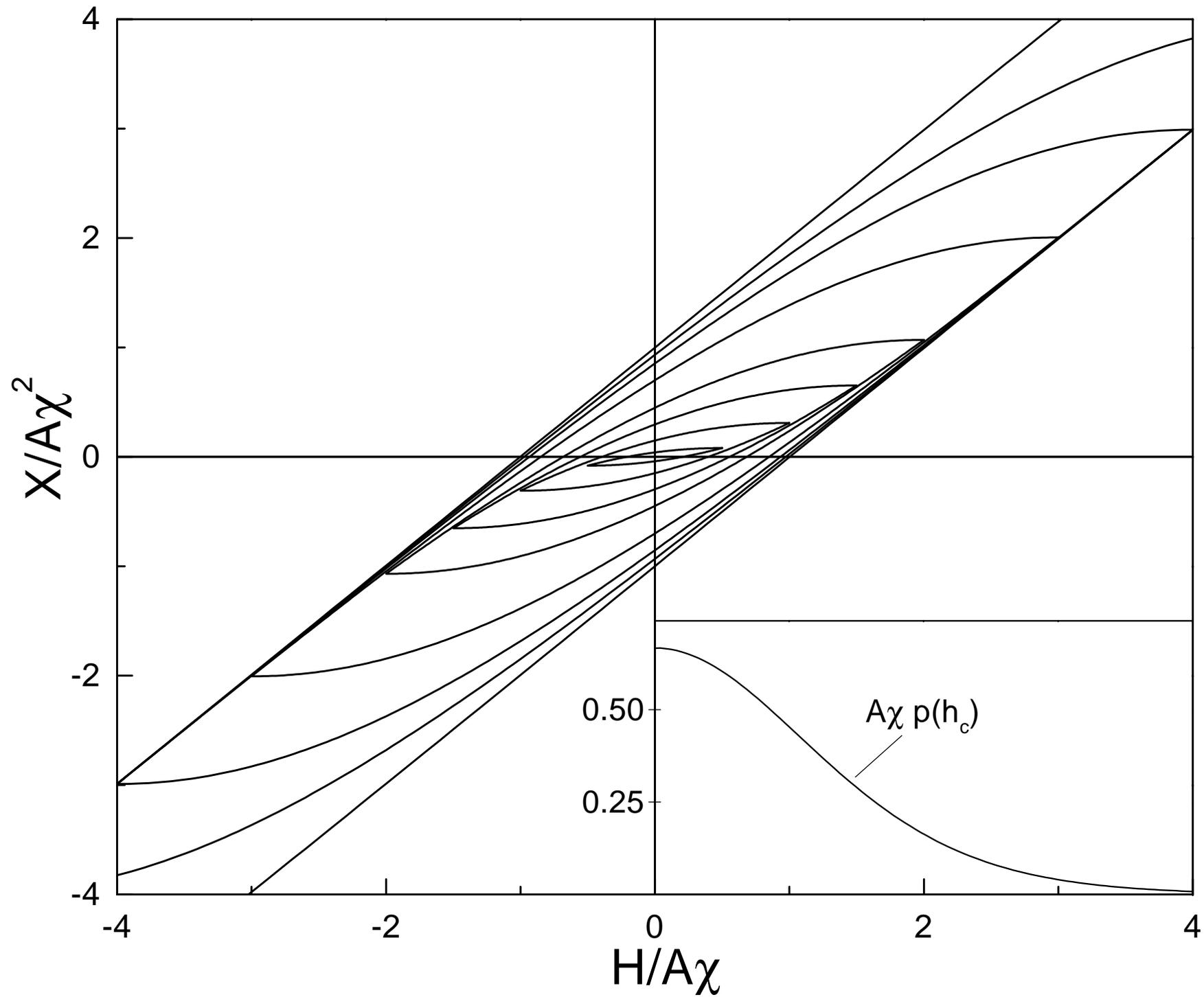